\documentstyle[12pt]{article}
\jot = 1.5ex\catcode`\@=11\renewcommand{\thefootnote}{\fnsymbol{footnote}}

\def\be{\begin{equation}}
\def\ee{\end{equation}}
\def\l{\label}
\def\F{{\cal F}}

\def\ldl{\Lambda\partial_\Lambda}

\begin{document}\begin{titlepage}

\vspace{1.cm}

\centerline{\large{\bf RG Flow Irreversibility, C-Theorem and}}

\vspace{.4cm}

\centerline{\large{\bf Topological Nature of 4D N=2 SYM}}

\vspace{1.5cm}

{\centerline{\sc Giulio BONELLI}}

\vspace{0.3cm}

\centerline{\it International School for Advanced Studies and INFN, 
Trieste, Italy}
\centerline{bonelli@sissa.it}

\vspace{1.cm}

{\centerline{\sc Marco MATONE}}

\vspace{0.3cm}

\centerline{\it Department of Physics ``G. Galilei'' and INFN, Padova, 
Italy}
\centerline{matone@padova.infn.it}

\vspace{2cm}

\centerline{\bf ABSTRACT}

\vspace{0.6cm}
\noindent
We determine the exact beta function and a RG flow
Lyapunov function
for $N=2$ SYM with gauge group $SU(n)$. It turns out that the {\it 
classical} discriminants of the Seiberg--Witten curves determine
the RG potential. The radial irreversibility of the RG flow in the $SU(2)$
case and the non--perturbative identity relating the $u$--modulus
and the superconformal anomaly,
indicate the existence of a four dimensional analogue of the  c--theorem
for $N=2$ SYM which we formulate for the full $SU(n)$ theory. 
Our investigation provides further evidence of the essentially topological 
nature of the theory.

\end{titlepage}

\newpage
\setcounter{footnote}{0}
\renewcommand{\thefootnote}{\arabic{footnote}}

Recently it has been shown that the 
exact results about $N=2$ SUSY Yang--Mills obtained by Seiberg and Witten 
\cite{SW1}  actually follow from first principles \cite{BMT}. 
In particular, in \cite{BMT} it has been shown
that the entire physical content of the $SU(2)$ theory can be 
extracted from the
identity \cite{mat}
\begin{equation}
u=\pi i ({\cal F}-a\partial_a{\cal F}/2).
\label{81}\end{equation}
In this context, we observe that
uniformization theory is the natural framework
for investigating $N=2$ SYM \cite{BMT}\cite{mat}.
A basic fact for the derivation in \cite{BMT} 
is that the identity (\ref{81}),
first checked up to 
two--instanton in \cite{FUTR}, has been proved to 
any order in the instanton expansion in \cite{DOKHMA} and has been
obtained as an anomalous superconformal
Ward identity in \cite{HW} (this also excludes other non--perturbative 
effects besides instantons).
A first consequence of these results is that 
$u=\langle Tr\phi^2\rangle$ is actually a good modular
invariant global coordinate \cite{mat}. 
In particular, one can consider the complex coupling constant
$\tau$ as a generally polymorphic
function of the independent variable $u\in {\bf C}\cup \{\infty\}$.
Furthermore, the $T^2$ symmetry $u(\tau+2)=u(\tau)$, 
which rigorously follows from the asymptotic analysis together with the 
relation (\ref{81}),
and the fact that
$\overline{u(\tau)}=u(-\bar\tau), u(\tau+1)=-u(\tau)$,
uniquely fix the monodromy group to be $\Gamma(2)$ \cite{BMT}
and therefore the explicit Seiberg--Witten results.

One of the main consequences of the Seiberg--Witten results is that
for the first time it has been possible to determine the exact expression
of the $\beta$--function  of
a non--trivial 
four dimensional quantum 
field theory \cite{Loro}\cite{Noi}. The exact expression
for the $\beta$--function in the case of $SU(3)$ was obtained
in \cite{Noi2}
(see also \cite{DKP} for related aspects). 
Very recently the $SU(2)$ $\beta$--function has been
reconsidered in a series of interesting papers
\cite{Essi1}\cite{Essi2}\cite{Essi3}. We will see
that the are non--trivial  structures which arise
in considering higher rank groups.

The exact solution for the $\beta$--function of the theory, provides the
possibility of looking for the analogue of the Zamolodchikov c--theorem
\cite{zam} in the context of four dimensional quantum field theories.
The content of the c--theorem is the identification of an RG monotonic
quantity, i.e. a Lyapunov function,  giving at its fixed points a way to
recognize some properties
of the conformal limits of the class of 2--D theories. As well known,
this quantity is strictly related to the conformal anomaly.
In the case of $2<D<4$ theories something can be guessed from the 
speculations in \cite{gaite} (and reference therein).
Our letter tries to give some contributions to the 4--D problem
(see for example \cite{varii} for related aspects).
In the case of the $SU(2)$ Seiberg--Witten theory, the results in
\cite{Essi2} can be understood from the c--theorem point of view, since 
(\ref{81}) means that 
$u$ is proportional to the (super)conformal anomaly \cite{HW}.

In this paper we show that this result fits in a more general framework in
which a Lyapunov function is naturally determined and
related to the {\it
classical} discriminant of the Seiberg--Witten curve.

Let us first consider some  aspect for the $SU(2)$ case.
We refer to \cite{BMT,mat,BS} for
the aspects related to uniformization theory.
Since the $u$ quantum moduli space ${\cal M}_{SU(2)}$
is the thrice punctured 
sphere, we have
$$u/\Lambda^2=J(\tau),$$
where $J$ is the uniformizing map 
$J:H\to {\bf C}\backslash\{\pm 1\}$
and $H$ is the upper half plane.

Since $J(i\infty)=\infty$, $J(\pm 1)=-1$, $J(0)=1$,  
the explicit expression of the $J$ map in terms of $\theta$--functions is
$$
J(\tau)=2{\theta_3^4\over \theta_2^4}-1,
$$
connected with the conventions of
\cite{Noi} by  $u\to -u$, $\theta_2\to \alpha \theta_3$,
$\theta_3\to \alpha \theta_2$,
$\theta_4\to \alpha^{-2}\theta_4$, $\alpha^4=-1$.

The exact $\beta$--function is \cite{Loro,Noi}
\be
\beta(\tau)=\Lambda{\partial\tau\over \partial \Lambda}|_u=-2 {J(\tau)\over 
J'(\tau)},
\l{betax}\ee
that in terms of $\theta$--functions has the form
\be
\beta(\tau)=-{i\over \pi}\left({1\over \theta_3^4}+{1\over \theta_4^4}\right).
\l{abbastanza}\ee
This expression
has been recently rederived in 
\cite{Essi1} and further investigated
in \cite{Essi2}\cite{Essi3} where it has been observed that by (\ref{betax})
\be
{d\tau\over\beta(\tau)}=\partial\Psi_2(\tau),
\l{daiiiije}\ee
where
\be
\Psi_2(\tau)=-{1\over 2}{\ln}|J|^2,
\l{erminioefisio}\ee
with $\partial=d\tau\partial_\tau$.

The radial irreversibility of the RG flow is proved just by noticing 
that
\be
\Lambda\partial_\Lambda|J|^2=
- 4 |J|^2,
\l{exenasarco}\ee
which means that $|u/\Lambda^2|^2$ is a non--increasing function along 
the RG flow. In other words 
\be
L_2= |J|^2=e^{-2\Psi_2},
\l{Lyapunov}\ee
is a Lyapunov function for the RG flow. Note that the only stable fixed 
point is $u=0$ which is ${\bf Z}_2$ invariant. It corresponds to the zero 
locus $\tau_0=\{\tau\in H|J(\tau)=0\}$, that is
$\tau_0=\left\{\gamma\cdot({i\pm 1\over 2})| \gamma\in 
\Gamma(2)\right\}$,
where $\Gamma(2)$ acts linearly fractionally on $\tau$.

It is clear that in view of the c--theorem, a basic step should be
to prove the existence of the potential for the $\beta$--function.
While in the $SU(2)$ case the derivation of this potential 
reduces to a simple integration, this is
a non--trivial task for higher rank groups.
We will see that the $\beta$--function potential exists
for $SU(n)$ for any $n\geq 2$. Actually, it turns out that the
structure introduced in \cite{Noi2} is the natural one 
to explicitly solve this problem.
As we will see, somewhat surprisingly,
 the potential is determined by the {\it classical}
discriminant of the Seiberg--Witten curves. 

In \cite{Noi2} it has been shown that the basic structures of the 
$SU(2)$ case  are naturally extended to $SU(3)$
if one introduces the modular invariant quantities
\begin{equation}
I^{\;\;  \gamma}_\beta=(\partial_k z)
{\left(\partial_\beta\tau\right)^{-1}}^{kl}
\partial_l u^\gamma, 
\label{hfgy}\end{equation}
where $\beta,\gamma=2,3$, $z$ is the modular invariant 
$z=a^k\partial_k\F-2\F$, $\partial_k=\partial_{a^k}$,
$\partial_\alpha=\partial_{u^\alpha}$ and $u^2\equiv u$, $u^3\equiv v$.
These expressions, which have been given in
\cite{Noi2} for $SU(3)$, trivially extend to $SU(n)$
for any $n\ge 2$.
For the modular invariant $z$ 
we have $z={3i\over \pi} u$, that is \cite{Noi2}
$u= {2\pi i\over 3}\left({\cal F}-{a^k\over 2}\partial_k\F\right)$,
which is the generalization of (\ref{81}) and has been derived by other
means and also for higher rank groups in \cite{STYEY}.

The above framework is the natural one  to properly investigating the
extension to $N=2$ SYM with higher rank gauge groups 
\cite{KLT}\cite{extension}.
For example, the modular invariant quantities
$I_\beta^{\;\;\gamma}$ allow us to find the analogue of the identity
(\ref{81}) in the case of $v$.

Let us consider the beta function (matrix)
\be
\beta_{ij}=\Lambda {\partial \tau_{ij}\over\partial 
\Lambda}|_{u^2,u^3,...}.
\l{iosq}\ee
Since under modular transformations 
$$
(a^D,a)\to ({a^D}',a')=
(Aa^D+Ba,Ca^D+Da), 
$$
$\left(\matrix{A\\ B\cr C\\D\cr}\right)\in Sp(2n-2,{\bf Z})$,
we have
$$
\beta\to(\tau C^t+D^t)^{-1}\beta(\tau C+D)^{-1},
$$
and 
$$
d\tau\to(\tau C^t+D^t)^{-1}d\tau(\tau C+D)^{-1}.
$$
It follows that
\be
b=\beta^{ij}d\tau_{ij}=\beta^{\alpha\delta}J_{\alpha\gamma\delta}
du^{\gamma},
\l{prepscro}\ee
is a modular invariant one--form.
Here $\beta^{ij}$ and $\beta^{\alpha\gamma}$ denote the inverse 
of the matrices $\beta_{ij}$ and $\beta_{\alpha\gamma}$
respectively, and
\be
J_{\alpha\beta\gamma}=\partial_\alpha a^i\partial_\beta \tau_{ij}
\partial_\gamma a^j,
\l{soiqh}\ee
are the modular invariant quantities
we introduced in \cite{Noi2}.
For $SU(3)$ these are related to the 
$I_\beta^{\,\,\gamma}$'s by
\be
I_\beta^{\,\,\gamma}J_{\gamma\beta2}={\pi\over 3i},\qquad
I_\beta^{\,\,\gamma}J_{\gamma\beta3}=0,
\l{odijdpl}\ee
whose solution is
\be
J_{222}=-{1\over 3}AP,\qquad J_{223}=12uvA,\qquad 
J_{233}=-{1\over u}AP,\qquad J_{333}=36vA,
\l{siq}\ee
where $A={3i\over \pi}[(12uv)^2-P^2/3u]^{-1}$ and
$P=27(v^2-\Lambda^6)+4u^3$. Since 
the explicit expression of $\beta_{\alpha\gamma}$ is \cite{Noi2}
\be
\beta_{22}={2Au\over 3}[P-54v^2],\qquad
\beta_{23}=\beta_{32}={3Av\over u}[P-8u^3],
\qquad
\beta_{33}=2A[P-54v^2],
\l{lmmsaf}\ee
it follows by
(\ref{prepscro}) and (\ref{siq}) that 
($\partial=du\partial_u+dv\partial_v$)
\be
b=\partial\Psi_3,
\l{daiee}\ee
where
\be
\Psi_3=-{1\over3}{ \ln}\left\vert{27v^2-4u^3\over 
\Lambda^6}\right\vert^2=
-{1\over3}{ \ln}\left\vert 27{\tt v}^2(\tau)-4{\tt u}^3(\tau)\right\vert^2,
\l{powplp}\ee
with 
${\tt u}(\tau)=u/\Lambda^2$, ${\tt v}(\tau)=v/\Lambda^3$.
Eq.(\ref{daiee}) shows that the RG flow is gradient.
Furthermore, as
\be
\Lambda\partial_\Lambda e^{-3\Psi_3}=
-12 e^{-3\Psi_3},
\l{dpr}\ee
it follows that 
\be
L_3= e^{-3\Psi_3},
\l{oicf}\ee 
is a Lyapunov function for the RG flow.

In \cite{Noi2} it has been observed that also in the $SU(3)$ case 
there is the uniformization mechanism which generalizes the
structure underlying the $SU(2)$ case \cite{mat}. In particular, 
the structure of the covering of the quantum moduli space 
${\cal M}_{SU(3)}$ is encoded in the properties of the Appell's functions. 
The fact that $\tau_{ij}$ is dimensionless implies that
\be
(\ldl + \Delta_{u,v})\tau_{ij}=0,
\l{oidx1}\ee
where $\Delta_{u,v}= 
2u\partial_u+3v\partial_v$ is the scaling invariant vector field. 
Eq.(\ref{oidx1}) implies that $\tau=\tau({\tt u},{\tt v})$.
Therefore, the $\tau$--space is a subvariety ${\cal S}$ of the genus 2 Siegel
upper--half space of complex codimension one covering ${\cal M}_{SU(3)}$.
In particular, 
the Picard--Fuchs equations (not to be confused with the reduced ones)
are the uniformizing equations for 
${\cal M}_{SU(3)}\cong {{\cal S}/M_{SU(3)}}$
where $M_{SU(3)}\subset Sp(4,{\bf Z})$ is the monodromy group
of the polymorphic matrix function $\tau$ seen as the inverse 
of the uniformizing map $\tau_{jk}=\tau_{jk}({\tt u},{\tt v})$.

For the higher rank case it is immediate to read out the general 
structure we are looking for. In fact our RG potentials $\Psi_2$ and 
$\Psi_3$ are simply related to the
{\it classical} discriminants of the Seiberg--Witten curves \cite{KLT}. 
These are defined as
$$
\Delta_{cl.}^{SU(n)}(u^\gamma)=\prod^n_{i<j}(e_i-e_j)^2,
$$
where $\{e_i\}$ are the zeros in $x$ of the polynomial ${\cal W}_{A_{n-1}}
(x;u^2,..,u^n)= 
x^n-\sum_{\gamma=2}^n u^\gamma x^{n-\gamma}$.
Explicitly, for $n=2,3$
$$
\Delta_{cl.}^{SU(2)}(u)=u,\qquad
\Delta_{cl.}^{SU(3)}(u,v)=4u^3-27v^2. 
$$
Therefore, there is strong evidence that for the $SU(n)$ case 
($\partial=\sum_{\gamma=2}^n d u^\gamma\partial_\gamma$)
\be
b=\partial\Psi_n,
\l{d1}\ee
where
\be
\Psi_n=-{1\over n}{\ln}\left\vert{ \Delta_{cl.}^{SU(n)}(u^\gamma)
\over\Lambda^{n(n-1)}}
\right\vert^2=
-{1\over n}{\ln}\left\vert \hat\Delta_{cl.}^{SU(n)}(\tau)
\right\vert^2,
\l{d2}\ee
with $b$ defined as in (\ref{prepscro}) and
$$
\hat\Delta_{cl.}^{SU(n)}(\tau)=\Delta_{cl.}^{SU(n)}({\tt u}^\gamma),
$$
where
$$
{\tt u}^\gamma=u^\gamma/\Lambda^\gamma, 
$$
$\gamma=2,\ldots,n$.
Furthermore, we have the equation
\be
\ldl L_n=-2n(n-1) L_n ,
\l{opjd}\ee
where
\be
L_n=
e^{-n\Psi_n}=\vert\hat\Delta_{cl.}^{SU(n)}(\tau)\vert^2.
\l{doiqj}\ee

Observe that by (\ref{prepscro}) it follows that
Eq.(\ref{d1}) is equivalent to
\be
\partial_\gamma \Delta_{cl.}^{SU(n)}(u^\sigma) =
\beta^{\alpha\delta}J_{\alpha\gamma\delta}\Delta_{cl.}^{SU(n)}(u^\sigma),
\l{FrankZappa}\ee
so that the integrability condition
\be
\partial_\gamma \partial_\sigma\Delta_{cl.}^{SU(n)}(u^\sigma)
=\partial_\sigma \partial_\gamma\Delta_{cl.}^{SU(n)}(u^\sigma),
\l{CC}\ee
yields
\be
\partial_\sigma (\beta^{\alpha\delta}J_{\alpha\gamma\delta})=
\partial_\gamma (\beta^{\alpha\delta}J_{\alpha\sigma\delta}),
\l{JimiHendrix}\ee
$\gamma,\sigma=1,\ldots, n-1$.

It is interesting to note that
the previous analysis implies the following 
scaling law for the classical discriminant
\be
\hat\Delta_{cl.}^{SU(n)}(\tau)=
\hat\Delta_{cl.}^{SU(n)}(\tau_0)e^{-n\int^\tau_{\tau_0}b},
\l{TT1}\ee
which is the higher rank version of the scaling law for
the $u$--modulus derived in \cite{Noi}. In this context we observe that
due to the presence of other moduli besides $u$, while in the $SU(2)$
case the scaling law (\ref{TT1}) implied, in view of (\ref{81}),
the RG equation for $\F$, this is no the case for
$SU(n)$, $n\geq 3$.

As $L_n$ reaches its minimum, we have 
$\hat\Delta_{cl.}^{SU(n)}(\tau)=0$, 
meaning that the system naturally tends to flow 
through the classical locus of gauge symmetry restoring: this restoring 
in fact does not really happen since the quantum moduli space is dramatically
different from the semiclassical one.
In any case, all this means that the classical symmetry restoring 
locus continues playing a non trivial attracting r\"ole
in the full theory.

Notice that the above result may cause some doubts: it is stated that the
exact quantum RG flow of a given theory follows some {\it classically} 
determined character. However observe that the classical character 
concerns the dependence of the potential on the quantum moduli
rather than the moduli themselves.
It seems that a full explanation
of the above phenomenon should be found in a deeper understanding of
non--renormalization theorems intertwined with the essentially topological
nature of the 
theory \cite{Witten}\cite{Noi2}\cite{MoWi}\cite{LoNeSh}\cite{bertmat}.

The fact that the theory has an essentially topological structure
has been suggested \cite{Noi2} where the WDVV equations \cite{WDVV}
for the SU(3) case have been derived in the framework
of the Picard--Fuchs equations. 
Using a different method, the WDVV equations 
for higher rank groups have
been obtained in \cite{MMM}.\footnote{The problem of deriving the 
WDVV equations for higher rank groups from 
the Picard--Fuchs equations has been recently solved in the
interesting paper by Ito and Yang \cite{ItoYang}.}

To show a possible connection with the WDVV equations,
we note that by (\ref{prepscro}) it follows that
\be
b=\beta^{ij}d\tau_{ij}=\beta^{ij}\F_{ijk}da^k,
\l{prepscroX}\ee
where 
$$
\F_{ijk}=\partial_i\partial_j\partial_k\F.
$$
Using $\sum_\alpha du^\alpha\partial_\alpha=\sum_kda^k\partial_k$, one has
that the integrability condition (\ref{CC}) implies
\be
(\F_{ijk}\partial_l-\F_{ijl}\partial_k)\beta^{ij}=0,
\l{CC3}\ee
$k,l=1,\ldots,n-1$. The point is that in \cite{Morozov} 
the WDVV equations have been obtained
as consistency condition for a system of differential equations
whose structure is reminiscent of Eq.(\ref{CC3}).
The appearance of the $\beta$--function 
suggests that Eq.(\ref{CC3}) corresponds to
the version of WDVV equations derived in \cite{bertmat} (see also 
\cite{Morozov,Carroll} for related aspects). 
The fact that the WDVV equations can be extended
by considering the RG scale $\Lambda$ as modulus 
\cite{bertmat2}, provides further evidence for the topological nature
of $N=2$ SYM (see also \cite{ItoYang} for related results).

A crucial point about our Lyapunov functions is whether they encode in some way
any physical information about the structure of the massless sector of the
theory at the critical points. 
We refer the reader to \cite{Essi3} and references therein for a
more general discussion about this aspect which is general enough to 
extend also to the higher rank case.

Finally, we observe that the approach in 
\cite{BS}\cite{IMNS}\cite{DKP} should be useful
to extend our results to the case with matter.
Another interesting aspect is that, as observed in
\cite{Essi3}, the
above structures are related
to the quantum Hall system \cite{LR} and non--linear sigma models \cite{HOS}.

\vspace{1cm}

\noindent
{\bf Acknowledgements}. It is a pleasure to thank
 G. Bertoldi, M. Bertolini, A. Cappelli, D.Z. Freedman,
J. Isidro, K. Konishi and M. Tonin for discussions.
MM was  supported in part by 
the European Commission TMR programme ERBFMRX--CT96--0045.

\end{document}